\documentclass[onecolumn,preprintnumbers,eqsecnum,superscriptaddress,10pt]{revtex4}
\usepackage{amsfonts}
\usepackage{amssymb}
\usepackage{amsmath}
\usepackage{epsf}
\usepackage{graphicx}
\usepackage{dcolumn}
\usepackage{bm}
\usepackage{CJK}
\usepackage{color}
\usepackage[colorlinks=true,
            citecolor=blue,
            urlcolor=blue,
            filecolor=black,
            linktocpage=true,
            linkcolor=blue,]{hyperref}

\def\nn{{\nonumber}}

\newcommand{\beq}{\begin{equation}}
\newcommand{\eeq}{\end{equation}}
\newcommand{\beqs}{\begin{eqnarray}}
\newcommand{\eeqs}{\end{eqnarray}}

\newcommand{\be}{\begin{equation}}
\newcommand{\ee}{\end{equation}}
\newcommand{\bea}{\begin{eqnarray}}
\newcommand{\eea}{\end{eqnarray}}

\def\Chone{C_{h1}}
\def\Chtwo{C_{h2}}

\def\Cktwo{C_{k2}}

\def\Cthree{C_{i1}}
\def\Cfour{C_{i2}}

\def\[{\left[}
\def\]{\right]}
\def\({\left(}
\def\){\right)}

\begin{document}

\title{Sonic velocity in holographic fluids and its applications}

\author{Ya-Peng Hu}\email{huyp@nuaa.edu.cn}
\address{College of Science, Nanjing University of Aeronautics and Astronautics, Nanjing 210016, China}
\address{Instituut-Lorentz for Theoretical Physics, Leiden University, Niels Bohrweg 2, Leiden 2333 CA, The Netherlands}
\address{Key Laboratory of Theoretical Physics, Institute of Theoretical Physics, Chinese Academy of Sciences, Beijing 100190, China}

\author{Yu Tian}\email{ytian@ucas.ac.cn}
\address{School of Physics, University of Chinese Academy of Sciences, Beijing 100190, China}

\author{Xiao-Ning Wu}\email{wuxn@amss.ac.cn}
\address{Institute of Mathematics, Academy of Mathematics and System Science, Chinese Academy of Sciences, Beijing 100190, China}
\address{Hua Loo-Keng Key Laboratory of Mathematics, CAS, Beijing 100190, China}

\author{Huai-Fan Li}\email{huaifan.li@stu.xjtu.edu.cn}
\address{Institute of Theoretical Physics, Department of Physics, Shanxi Datong University, \\Datong 037009, China}

\author{Hongsheng Zhang }\email{sps\_zhanghs@ujn.edu.cn}
\address{School of Physics and Technology, University of Jinan, 336 West Road of Nan Xinzhuang, Jinan, Shandong 250022, China}

\begin{abstract}
   Gravity/fluid correspondence becomes an important tool to investigate the strongly correlated fluids. We carefully investigate the holographic fluids at the finite cutoff surface by considering different boundary conditions in the scenario of gravity/fluid correspondence. We find that the sonic velocity of the boundary fluids at the finite cutoff surface is critical to clarify the superficial similarity between bulk viscosity and perturbation of the pressure for the holographic fluid, where we set a special boundary condition at the finite cutoff surface to explicitly express this superficial similarity. Moreover, we further take the sonic velocity into account to investigate a case with more general boundary condition. In this more general case, two parameters in the first order stress tensor of holographic fluid cannot be fixed, one can still extract the information of transport coefficients by considering the sonic velocity seriously.


\end{abstract}

\maketitle



\section{Introduction}

The AdS/CFT
correspondence~\cite{Maldacena:1997re,Gubser:1998bc,Witten:1998qj,Aharony:1999ti}
 is a significant progress in theoretical physics. This correspondence  provides
 new insights and useful tools to investigate the strongly related field theory by using the weakly correlated gravity
 theory~\cite{Zaanen:2015oix,Herzog:2009xv,CasalderreySolana:2011us,Policastro:2001yc,Buchel:2003tz,Kovtun:2004de}.

 At long wave limit the AdS/CFT correspondence reduces to gravity/fluid correspondence \cite{Bhattacharyya:2008jc}. In the gravity/fluid correspondence, the dual field theory
usually resides on the infinite boundary (conformal boundary or UV boundary), and has conformal dynamics~\cite{Bhattacharyya:2008jc,Hur:2008tq, Erdmenger:2008rm, Banerjee:2008th,
Son:2009tf, Tan:2009yg, Torabian:2009qk, Hu:2010sn, Hu:2011ze,
Kalaydzhyan:2011vx, Amado:2011zx}. In fact, the gravity/fluid correspondence can
be generalized to study non-conformal dual systems. A simple way of achieving
this is to break the conformal symmetry by introducing a finite
cutoff on the radial coordinate in the bulk, which has implied a deep relation between the Navier-Stokes (NS)
equations and the Einstein
equations~\cite{Bredberg:2010ky,Bredberg:2011jq,Compere:2011dx,Cai:2011xv,Niu:2011gu,Kuperstein:2011fn,Compere:2012mt,Kuperstein:2013hqa}. In addition, from the renormalization group (RG) viewpoint, the radial
direction of the bulk spacetime corresponds to the energy scale of
the dual field theory~\cite{Balasubramanian:1998de,Akhmedov:1998vf,de
Boer:1999xf,Susskind:1998dq,Bredberg:2010ky,Heemskerk:2010hk,Faulkner:2010jy,Iqbal:2008by,Sin:2011yh}. Thus investigations to holographic fluids at a finite cutoff surface have started\cite{Bredberg:2010ky,Niu:2011gu,Cai:2011xv,Kuperstein:2011fn,Bai:2012ci,Hu:2013lua,Brattan:2011my,Camps:2010br,Emparan:2012be,Emparan:2013ila}.
Generally the holographic fluid on the cutoff surface is usually non-conformal~\cite{Brattan:2011my,Camps:2010br,Emparan:2012be,Emparan:2013ila,Bai:2012ci,Hu:2013lua}.

In this paper, we focus on the stress tensor of non-conformal fluids, whose transport coefficients are obtained from the holography.  It is found that the sonic velocity of the holographic fluids at the finite cutoff surface is critical to clarify the superficial similarity between bulk viscosity and perturbation of the pressure for holographic fluids, and further simplifies the first order stress tensor of holographic fluid at the finite cutoff surface. Under more general boundary conditions at the finite cutoff surface, we also investigate applications of sonic velocity of non-conformal holographic fluids in details in this paper.

This article is organized as follows. In Sec.~II, we focus on the first order perturbative solution of the Schwarzschild-AdS black brane solution. Since this part is simple and fundamental which can be seen in the previous works, here we just give a brief review as a warmup to make the whole paper more readable. In Sec.~III, several boundary conditions are carefully analyzed, which is classified into two cases under the choice of boundary condition $h(r_c)$. Besides the general expressions of perturbations of pressure and energy density in the renormalization group (RG) the holographic fluid are expressed, the superficial similarity has been also explicitly seen between the bulk viscosity and perturbation of pressure. A most important result is that we propose a method to distinguish this superficial similarity through studies of the sonic velocity in the holographic fluid. Moreover, we further take the sonic velocity into account to investigate a more general boundary condition case with $h(r_c)\neq0$. Sec.~IV is devoted to the conclusion and discussion. Note that, Latin index repeated is usually represented to take the summation in our whole paper. For example, $\partial_i \beta_i=\Sigma_{i}\partial_i \beta_i$. However, if it is not so, it will be pointed out in equations, et al.

\section{Warmup: The first order perturbative solution of the Schwarzschild-AdS black brane solution}
We make a concise review of the Schwarzschild-AdS black brane. The action of five-dimensional Einstein gravity with a negative cosmological constant $\Lambda=-6/\ell^2$ reads
\begin{equation}
\label{IVaction1} I=\frac{1}{16 \pi G}\int_\mathcal{M}~d^5x
\sqrt{-g^{(5)}} \left(R-2 \Lambda
\right).
\end{equation}
The corresponding field equation reads
\begin{eqnarray}
\label{IVeqs1}
R_{AB } -\frac{1}{2}Rg_{AB}+\Lambda g_{AB}&=&0~.
\end{eqnarray}
Here the AdS radius $\ell=1$ and $16\pi G=1$ have been set for later convenience. The Schwarzschild-AdS black brane solution is
\begin{eqnarray}
ds^2=\frac{dr^2}{r^2f(r)}+r^2
 \left(\mathop\sum_{i=1}^{3}dx_i^2 \right)-r^2f(r) dt^2, \label{IVSolution}
\end{eqnarray}%
where
\begin{eqnarray}
\label{IVf-BH} f(r) &=& 1-\frac{2M}{r^{4}},
\end{eqnarray}%
The Hawking temperature of the Schwarzschild-AdS black brane solution is
\begin{eqnarray}
T_{+}&=&\frac{(r^2f(r))'}{4 \pi}|_{r=r_{+}}=\frac{r_{+}}{\pi}\ ,\label{IVTemperature}
\end{eqnarray}
where $r_{+}$ is the location of horizon and positive root of $f(r)=0$.

In the Eddington-Finkelstin coordinates, the black brane becomes
\begin{eqnarray}\label{IVSolution1}
ds^2 &=& - r^2 f(r)dv^2 + 2 dv dr + r^2(dx^2 +dy^2 +dz^2),
\end{eqnarray}
where $v=t+r_*$, and $r_*$ is the tortoise coordinate satisfying
$dr_*=dr/(r^2f)$. Note that, the holographic fluid is investigated to reside at some cutoff
hypersurface with constant radial coordinate $r=r_c$ ($r_c$ is a constant). It is helpful to
 make the following coordinates transformation $v\rightarrow v/\sqrt{r_c^2 f(r_c)}$ and $x_i\rightarrow x_i/r_c$ in
the solution (\ref{IVSolution1}), which makes the
induced metric on the cutoff surface to be explicitly flat metric, i.e. the cutoff surface with metric
$ds^2=-dv^2+dx^2+dy^2 +dz^2$.  The Hawking temperature is expressed as $T=T_{+}/\sqrt{r_c^2 f(r_c)}$ with respect to the killing observer $(\partial/\partial v)^a$ in the new
coordinate system, and the Schwarzschild-AdS black brane solution becomes
\begin{eqnarray}\label{IVSolution2}
ds^2 &=& - \frac{r^2 f(r)}{r_c^2 f(r_c)}dv^2 + \frac{2}{r_c \sqrt{f(r_c)}} dv dr + \frac{r^2}{r_c^2}(dx^2 +dy^2 +dz^2),
\end{eqnarray}
while the entropy density is $s=\frac{r_+^3}{4 G r_c^3}$. The boosted Schwarzschild-AdS black brane solution is
\begin{eqnarray}   \label{IVrnboost}
ds^2 &=& - \frac{r^2 f(r)}{r_c^2 f(r_c)}( u_\mu dx^\mu )^2 - \frac{2}{r_c \sqrt{f(r_c)}} u_\mu dx^\mu dr + \frac{r^2}{r_c^2} P_{\mu \nu} dx^\mu dx^\nu,  ~~
\end{eqnarray}
with
\begin{equation}
u^v = \frac{1}{ \sqrt{1 - \beta_i^2} },~~u^i = \frac{\beta_i}{
\sqrt{1 - \beta_i^2} },~~P_{\mu \nu}= \eta_{\mu\nu} + u_\mu u_\nu\ ,
\label{IVvelocity}
\end{equation}
where $x^\mu=(v,x_{i})$ is the boundary coordinates at the cutoff surface, $P_{\mu \nu}$ is the projector onto spatial directions, velocities $\beta^i $ are constants, and the boundary indices $(\mu,\nu)$ are raised and lowered by using the Minkowski
metric $\eta_{\mu\nu}$, while the bulk indices are distinguished by $(A,B)$.

We define a useful tensor
\begin{eqnarray}
&&W_{AB} = R_{AB} + 4g_{AB}, \label{IVTensors1}
\end{eqnarray}
while solutions of equation motions are equivalent to $W_{AB}=0$. Viewed from the gravity/fluid correspondence scenario, one needs perturb the gravitational solutions in the bulk spacetime to obtain transport coefficients of holographic fluids like shear viscosity $\eta$. The general procedure is promoting the constant parameters $\beta^i $ and $M$ in (\ref{IVrnboost}) to functions of boundary
coordinates $x^\mu$, i.e. $\beta^i (x^\mu)$ and $M(x^\mu)$ ~\cite{Bhattacharyya:2008jc,Hur:2008tq}. Therefore, ~(\ref{IVrnboost}) will be no longer the solution of the field equation (\ref{IVeqs1}) since the parameters now depend on the boundary coordinates, and hence extra correction terms are
needed to add to make (\ref{IVrnboost}) be a self-consistent solution.

For the extra correction terms, we can just focus on the extra correction terms around the origin $x^{\mu}=0$, and the first order extra correction terms around $x^\mu=0$ are \cite{Bhattacharyya:2008jc}
\begin{eqnarray}\label{IVcorrection}
&&ds_{(1)}^2 = \frac{ k(r)}{r^2}dv^2 +
2\frac{h(r)}{r_c \sqrt{f(r_c)}}dv dr + 2 \frac{j_i(r)}{r^2}dv dx^i
 +\frac{r^2}{r_c^2} \left(\alpha_{ij}(r) -\frac{2}{3} h(r)\delta_{ij}\right)dx^i dx^j,
\end{eqnarray}
where an appropriate gauge has been chosen, i.e. the background field gauge in \cite{Bhattacharyya:2008jc} ($G_{AB}$
represents the full metric)
\begin{equation}
G_{rr}=0,~~G_{r\mu}\propto u_{\mu},~~Tr((G^{(0)})^{-1}G^{(1)})=0,\label{gauge}
\end{equation}
while $G^{(0)}$, $G^{(1)}$ are the corresponding zero order and first order terms in $G_{AB}$, and $\alpha_{ij}(r)$ is in fact traceless for this background field gauge since $Tr((G^{(0)})^{-1}G^{(1)})=\mathop\sum_{i}\alpha_{ii}$. Note that, parameters around $x^\mu=0$ expanded to the first order are
\begin{eqnarray}
\beta_i(x^\mu)&=&\partial_{\mu} \beta_{i}|_{x^\mu=0}
x^{\mu},~M(x^\mu)=M(0)+\partial_{\mu}
 M|_{x^\mu=0} x^{\mu}, \label{IVExpand}
\end{eqnarray}
where $\beta_i(0)=0$ are assumed at the origin $x^{\mu}=0$. Thus after inserting the metric (\ref{IVrnboost}) with non-constant parameters and (\ref{IVExpand}) into $W_{AB }$ , the nonzero $-W_{AB }$ is usually considered as the first order source terms $S^{(1)}_{AB }$, while the first order perturbation solution around $x^\mu=0$ can be obtained
 from the vanishing $W_{AB} = (\text{effect from correction})
  - S^{(1)}_{AB }$, which are casted into the appendix~\ref{A}.

 Still, there are two constraint equations
\begin{eqnarray}
 &&W_{vv} + \dfrac{r^2 f(r)}{ r_c\sqrt{f(r_c)}}W_{vr} =0 ~\Rightarrow~ S_{vv}^{(1)} + \dfrac{r^2 f(r)}{r_c\sqrt{f(r_c)}}S_{vr}^{(1)} = 0, \notag \\
 &&W_{vi} + \dfrac{r^2 f(r)}{r_c\sqrt{f(r_c)}} W_{ri} =0 ~\Rightarrow~ S_{vi}^{(1)} +\dfrac{r^2 f(r)}{r_c\sqrt{f(r_c)}}S_{ri}^{(1)} = 0. \label{constraint}
 \end{eqnarray}
From the appendix~\ref{A}, one rewrites these constrain equations~(\ref{constraint}) as
\begin{eqnarray}
&&3 \partial _vM+4 M \partial _i\beta _i=0,\\\nn
&& \partial _iM+4 M \partial _v\beta _i=\frac{- 4 M \partial _i M }{r_c^4 f\left(r_c\right)}~~,
\end{eqnarray}
which are nothing but the conservation equations of the zeroth order stress-energy tensor~\cite{Bhattacharyya:2008jc, Hur:2008tq,Hu:2010sn, Hu:2011ze}. Further, one analytically
obtains
\begin{eqnarray}
h(r)&=&{\Chtwo}+\frac{{\Chone}}{r^4},\nn\\
k(r)&=&{\Cktwo}-\frac{2 {\Chtwo} r^4}{r_c^2 f\left(r_c\right)}+\frac{4 {\Chone} M}{3 r^4 r_c^2 f\left(r_c\right)}+\frac{2 r^3 \partial _i\beta _i}{3 r_c \sqrt{f\left(r_c\right)}},\nn\\
j_i(r)&=&\frac{r^3}{r_c^5f(r_c)^{\frac{3}{2}}}\left(\partial _i M+r_c^4 f(r_c)\partial_v \beta_i\right)+\frac{C_{i1} r^4}{4}+C_{i2} \nn\\
&=&\frac{r^3 r_c^3 \sqrt{f\left(r_c\right)}}{2M+r_c^4} \partial_v \beta_i+\frac{C_{i1} r^4}{4}+C_{i2},\label{Coefficient}\nn\\
\alpha_{ij}(r)&=&\alpha(r)\left\{(\partial_i \beta_j + \partial_j
\beta_i )-\frac{2}{3} \delta_{ij}\partial_k \beta^k \right\}, \label{alpha_ij}
\end{eqnarray}
where $\alpha(r)$ is $\alpha(r)= r_c\sqrt{f(r_c)} \int_{r_c }^{r}\frac{s^{3}-r_{+}^3}{-s^{5}f(s)}ds$, and $\Chone, \Chtwo, \Cktwo, \Cthree$ and $\Cfour$ are nine constants of integration.

\section{The stress tensor of first order holographic fluid under different boundary conditions at the finite cutoff surface  }

Note that, the previous works usually investigate the holographic fluid just residing at the UV boundary or infinite cutoff surface (i.e., $r_c$ to infinity)~\cite{Bhattacharyya:2008jc,Hur:2008tq}. Here we will have a try to use the gravity/fluid correspondence to shed some insights on the holographic fluid at the finite cutoff surface, which can be considered as a simple generalization of the previous works. However, it should be emphasized that this generalization is non-trivial, since the stress tensor of the holographic fluid at the finite cutoff surface is usually non-conformal and depends on the choice of boundary conditions. All of those points can be seen more clearly in the following contents.

According to the gravity/fluid correspondence, the stress tensor $T_{\mu\nu}$ of holographic fluid residing at the cutoff surface with the induced metric $\gamma_{\mu\nu}$ is given by~\cite{Bredberg:2010ky,Bredberg:2011jq,Myers:1999psa,Balasubramanian:1999re,de
Haro:2000xn,Emparan:1999pm,Mann:1999pc}
\begin{equation}
 T_{\mu\nu}=2\left(K_{\mu\nu}-K\gamma _{\mu\nu}-C\gamma _{\mu\nu}\right)\ , \label{IVTabCFT}
\end{equation}
where $\gamma_{\mu\nu}$ is the boundary metric obtained from the usual ADM decomposition
\begin{eqnarray}
ds^2 = \gamma_{\mu\nu}(dx^\mu + V^\mu dr)(dx^\nu + V^\nu dr) + N^2
dr^2\ ,
\end{eqnarray}
the extrinsic curvature is $K_{\mu\nu}=-\frac{1}{2}(\nabla_{\mu}n_{\nu}+\nabla_{\nu}n_{\mu})$, and $n^{\mu}$ is the unit normal vector of the constant hypersurface
$r=r_c$ pointing toward the $r$ increasing direction. In addition, the term $C\gamma _{\mu\nu}$ is usually related to the boundary counterterm added to cancel the divergence of the stress tensor $T_{\mu\nu}$ when the boundary $r=r_c$ approaches to infinity, for
example $C=3$ in the asymptotical $AdS_5$ case. However, there is no divergence of the stress tensor in our case with finite boundary. In the following, we still add the boundary counterterm with $C=3$ in the stress tensor, and the simple reason is that we require that our result should reduce to the previous result when $r_c$ goes to infinity ~\cite{Bhattacharyya:2008jc,Hur:2008tq,Bai:2012ci,Hu:2013lua}. Therefore, after obtaining the first order perturbative solution in the bulk, we can obtain the general formula of stress tensor $T_{\mu\nu}$ of the holographic fluid at the cutoff surface, i.e. around the origin $x^{\mu}=0$
\beqs
    T^{(0)}_{v v} &=& 2\left(C-3 \sqrt{f\left(r_c\right)}\right),\nn\\
    T^{(0)}_{x x} &=&T^{(0)}_{y y} =T^{(0)}_{z z} =\frac{-4 M+2\left(3-C \sqrt{f\left(r_c\right)}\right) r_c^4}{\sqrt{f\left(r_c\right)} r_c^4}, \label{STBackground}\\
    T^{(1)}_{v v} &=&-2\partial _i\beta _i +6 \sqrt{f\left(r_c\right)}h\left(r_c\right)+\frac{\left(-2 C+9 \sqrt{f\left(r_c\right)}\right) k\left(r_c\right)}{ r_c^2}+2\sqrt{f\left(r_c\right)} r_c h'\left(r_c\right),\nn\\
    T^{(1)}_{v i} &=& \frac{\partial _iM}{f\left(r_c\right) r_c^4}-\partial _v\beta _i +2\frac{\left(2-C \sqrt{f\left(r_c\right)}+3 f\left(r_c\right)\right)  j_i\left(r_c\right)}{\sqrt{f\left(r_c\right)} r_c^2}-\frac{\sqrt{f\left(r_c\right)} j_i'\left(r_c\right)}{ r_c},\nn\\
    T^{(1)}_{i j} &=&2\left( \delta _{i j}\partial _k\beta _k-\partial _{(i}\beta _{j)} \right)+2\delta _{i j}\frac{\partial_v M}{f\left(r_c\right) r_c^4}
 +2\left(-C+\frac{-2 M+3 r_c^4}{\sqrt{f\left(r_c\right)} r_c^4}\right)a_{i j}\left(r_c\right)-\sqrt{f\left(r_c\right)} r_c a_{i j}'\left(r_c\right)\nn\\
    & & +2\delta _{i j}\left(\left(\frac{2 C }{3}+\frac{5 \left(2 M-3 r_c^4\right)}{3 \sqrt{f\left(r_c\right)} r_c^4}\right) h\left(r_c\right)-\frac{2}{3} \sqrt{f\left(r_c\right)} r_c h'\left(r_c\right)\right)\nn\\
    & & +2\delta _{i j}\left(\frac{\left(-2 M+\left(3-2 f\left(r_c\right)\right) r_c^4\right) k\left(r_c\right)}{2 \sqrt{f\left(r_c\right)} r_c^6}-\frac{\sqrt{f\left(r_c\right)} k'\left(r_c\right)}{2
    r_c}\right).\label{BoundaryST}
\eeqs
Obviously, the further explicit results of first order stress tensor depend on several conditions, and hence extract the information of transport coefficients. In the following, we will carefully investigate the boundary conditions, particularly the boundary condition related to $h(r_c)$, because the cases under this boundary condition choice are complicate. Moreover, in fact this boundary condition can be relaxed to arbitrary at the finite cutoff surface, which has not been investigated before.

\subsection{Boundary condition with $h(r_c)=0$}
It is clear that one can fix the nine
parameters $\Chone, \Chtwo, \Cktwo, \Cthree$ and $\Cfour$ in (\ref{Coefficient}) to extract the exact transport coefficients of first order holographic fluid at the finite cutoff surface in (\ref{BoundaryST}). Therefore, several conditions can be assumed. In fact, the Dirichlet boundary condition is usually chosen in (\ref{IVcorrection}) like~\cite{Brattan:2011my,Bai:2012ci,Hu:2013lua,Cai:2011xv}
\beqs
    h(r_c) = 0 ,\quad k(r_c) = 0,\quad j_i(r_c) = 0.~\label{DirichletB}
\eeqs
In addition, another condition can be also chosen
\begin{equation}
T^{(1)}_{v i} =0,~\label{LandauFrame}
\end{equation}
since $T^{(1)}_{v i} =0$ is a gauge choice usually considered in the Landau frame, i.e. $T^{(1)}_{v v}=T^{(1)}_{v i}=0$ which corresponds that the velocity $u^{\mu}$ is identified as the 4-velocity of energy of relativistic fluid or a (normalized) timelike eigenvector of $T_{\mu\nu}$. Therefore, one final condition is needed to fix the nine parameters.
Note that, obviously, the final condition can be chosen as $T^{(1)}_{v v}=0$, which is just the Landau frame case with (\ref{LandauFrame}), and the corresponding results have been explicitly obtained in the appendix~\ref{B}. However, from (\ref{BoundaryST}), we find that $T^{(1)}_{v v}=0$ under (\ref{DirichletB}) just corresponds some special boundary condition related to $h'(r_c)$, while $T^{(1)}_{v v}$ will be non-zero for many other boundary condition cases, i.e. $h'(r_c)=0$. Therefore, it will be interesting to investigate another special boundary condition case, i.e., $h(r_c)=0$ and $h'(r_c)$ is kept as an arbitrary constant. Moreover, one will find that this special boundary condition will be also critical to explicitly see the superficial similarity between the bulk viscosity and perturbation of pressure in the stress tensor of the holographic fluid, while $T^{(1)}_{v v}=0$ case is a little more difficult to see this superficial similarity. Therefore, in the following, we will just focus on carefully investigating the stress tensor of holographic fluid under this special boundary conditions case.

From (\ref{Coefficient}), it is easy to find that keeping $h'(r_c)$ as an arbitrary constant is equivalent to keep the parameter $C_{h1}$ as an arbitrary constant. Therefore, the other eight parameters $\Chtwo, \Cktwo, \Cthree$ and $\Cfour$ can be solved from (\ref{DirichletB}) and (\ref{LandauFrame}), which are all expressed in $C_{h1}$
\begin{eqnarray}
&&{\Chtwo}=-\frac{C_{h1}}{r_c^4},~~{\Cktwo}=-\frac{2\partial _i\beta _i
r_c^2}{3\sqrt{f\left(r_c\right)}}-\frac{2C_{h1}(2 M+3 r_c^4)}{3r_c^6f(r_c)},\nn\\
&&{\Cthree}=-\frac{4 r_c^2 \partial _v\beta _i}{\sqrt{f(r_c)}(2M+r_c^4)},\quad{\Cfour}=\frac{2 M r_c^2 \partial _v\beta _i}{\sqrt{f(r_c)}(2M+r_c^4)}.
\label{NewParameters}
\end{eqnarray}
After inserting (\ref{NewParameters}) into (\ref{BoundaryST}), the non-zero components of stress tensor $T^{(1)}_{\mu\nu}$
are \beqs
    & &T^{(1)}_{v v}=-2\partial _i\beta _i +2 r_c \sqrt{f(r_c)} h'(r_c)=-2\partial _i\beta _i -\frac{8 \sqrt{f(r_c)}C_{h1}}{r_c^4},\nn\\
   & &T^{(1)}_{i j} = \frac{- 2r_+^3\sigma _{ij}}{r_c^3}+\delta_{ij}\big(\frac{-2(2 M +r_c^4)}{3( -2 M+r_c^4)}\partial _k\beta _k-\frac{8(2 M+r_c^4)C_{h1}}{3r_c^8\sqrt{f(r_c)}}). \label{BoundaryST1}
\eeqs
Note that, if the fluid is not considered under the Landau frame, usually the stress tensor of holographic fluid at the cutoff surface with the induced metric $\gamma _{\mu\nu}=\eta _{\mu\nu}$ can be written in a general form~\cite{Emparan:2013ila}
\begin{eqnarray}
T_{\mu \nu}=\rho\,u_\mu u_\nu+p
P_{\mu\nu}-2\eta\sigma_{\mu\nu}-\zeta\theta P_{\mu\nu} -\zeta^{'}\theta u_\mu u_\nu-\kappa a_{(\mu}u_{\nu)},
\label{StressTensor1}
\end{eqnarray}
where
\begin{eqnarray}
P_{\mu\nu}&=&\eta_{\mu\nu}+u_{\mu}u_{\nu},~
\sigma ^{\mu  \nu }\equiv \frac{1}{2} P^{\mu  \alpha } P^{\nu
\beta } \left(\nabla _{\alpha }u_{\beta } +\nabla _{\beta }u_{\alpha
}\right)-\frac{1}{3} P^{\mu  \nu } \nabla _{\alpha }u^{\alpha },~\theta=\nabla_{\mu} u^{\mu},~a^{\nu}=u^{\mu}\nabla_{\mu} u^{\nu},
\end{eqnarray}
and $\zeta^{'}$ is a shift of the local energy density by the expansion of the fluid, while $\kappa$ is the heat conductivity. In our case, if we still consider the fluid with the velocity in (\ref{IVvelocity}), the above form of stress tensor can be
\begin{eqnarray}
T_{\mu \nu}=\rho\,u_\mu u_\nu+p P_{\mu\nu}-2\eta\sigma_{\mu\nu}-\zeta\theta P_{\mu\nu},
\label{StressTensor2}
\end{eqnarray}
where $a^{v}=0,~a^{i}=\partial_v \beta_i$ around the $x^{\mu}=0$ has been used in our case and the $T^{(1)}_{v i}\neq 0$ can be cancelled by the gauge choice in (\ref{LandauFrame}), and it should be pointed out that here $\rho$ and $p$ can contain the first order terms with respect to the derivative of velocity although the stress tensor form looks like the form under the Landau frame.

After the comparison between the results in (\ref{BoundaryST1}) with (\ref{StressTensor2}), it will be easy to identify the energy density $\rho$ and shear viscosity $\eta$. However, a superficial similarity between the pressure $p$ and bulk viscosity $\zeta$ is explicitly seen in this case. Note that, from (\ref{STBackground}), the zero order pressure and energy density of dual fluid are $p_0=\frac{-4 M+2\left(3-3\sqrt{f\left(r_c\right)}\right)r_c^4}{r_c^4\sqrt{f\left(r_c\right)}}$, $\rho_0=2\left(3-3 \sqrt{f\left(r_c\right)} \right)$, and hence the entropy density $s$ of dual fluid can be computed through
\begin{eqnarray}
s=\frac{\partial p_0}{\partial T}=4\pi \frac{r_+^3}{r_c^3},
\end{eqnarray}
which is consistent with the entropy density of the black brane solution (\ref{IVSolution2}) with $16\pi G=1$ recovered, and it is convenient to check this equation if we express $p_0$ and $T$ in the functions of $r_+$. Furthermore, it can be easily checked that the familiar thermodynamic relation still holds on arbitrary cutoff surface for the zero order pressure and energy density
\begin{eqnarray}
\rho_{0}+p_0-Ts=0,
\end{eqnarray}
where $T$ is the temperature of the dual fluid related to the Hawking temperature of the black brane solution by $T=T_{+}/\sqrt{r_c^2 f(r_c)}$. Therefore, the precise underlying superficial similarity is in fact between the perturbation of pressure $p$ and the bulk viscosity $\zeta$, i.e., the term proportional to $\partial _k\beta _k$ in $T^{(1)}_{i j}$ in the (\ref{BoundaryST1}) belongs to the perturbation of pressure or the bulk viscosity.  For example, there can be two simple different choices, the first choice is
\begin{eqnarray}
&&\rho=2\left(3-3 \sqrt{f\left(r_c\right)} \right) -2\theta-\frac{8 \sqrt{f(r_c)}C_{h1}}{r_c^4},~~\eta=\frac{r_+^3}{r_c^3},\nn\\
&&p=\frac{-4 M+2\left(3-
3\sqrt{f\left(r_c\right)}\right)
r_c^4}{r_c^4\sqrt{f\left(r_c\right)}}-\frac{8(2 M+r_c^4)C_{h1}}{3r_c^8\sqrt{f(r_c)}},~~\zeta=\frac{2(2 M +r_c^4)}{3( -2 M+r_c^4)}, \label{ets}
\end{eqnarray}
while the other is
\begin{eqnarray}
&&\rho=2\left(3-3 \sqrt{f\left(r_c\right)} \right) -2\theta-\frac{8 \sqrt{f(r_c)}C_{h1}}{r_c^4},~~\eta=\frac{r_+^3}{r_c^3},\nn\\
&&p=\frac{-4 M+2\left(3-
3\sqrt{f\left(r_c\right)}\right)
r_c^4}{r_c^4\sqrt{f\left(r_c\right)}}-\frac{8(2 M+r_c^4)C_{h1}}{3r_c^8\sqrt{f(r_c)}}-\frac{2(2 M +r_c^4)}{3( -2 M+r_c^4)}\theta,~~\zeta=0. \label{ets1}
\end{eqnarray}
However, (\ref{ets}) and (\ref{ets1}) cannot satisfy at the
same time the thermodynamic relation between energy density and
pressure. In addition, the bulk viscosity should be only one number in the same boundary condition case. Moreover, the bulk viscosity can increase the total entropy of fluid, and hence usually it is different from the other pressure term although sometimes it is also considered as the effective pressure. Therefore, we should use an underlying method to extract the physical information of the holographic fluids. In fact, after making a careful consideration, we will find that there are two subtleties in the first choice or consideration (\ref{ets}). First, the $T^{(1)}_{v v}=0$ case as a special case contained in (\ref{BoundaryST1}) has been explicitly shown in the appendix~\ref{B}, and the bulk viscosity is zero, which will not be consistent with the results in the first choice with a nonzero bulk viscosity in (\ref{ets}). Second, the $C_{h1}$ term in (\ref{ets}) can be also considered as the bulk viscosity term, particularly when it is also proportional to $\partial _k\beta _k$ in some boundary condition case, and hence in fact there is also an underling ambiguity for the choice of bulk viscosity related to the term $C_{h1}$ in (\ref{ets}). Therefore, for further obtaining the true transport coefficients particular the bulk viscosity, one needs find out a method.

In the following, we will propose a method by checking the underlying consistency in (\ref{ets}) or (\ref{ets1}) with the thermodynamic relation between energy density and
pressure, i.e. through the studies of sonic velocity $c_s$ between the perturbations of energy density and pressure. As we know, the first order term in $\rho$ in fact can be also considered as the perturbation of energy density $\delta\rho$, while this perturbation of energy density usually deduces the perturbation of pressure of fluid $\delta p$. In our case, using the above explicit expressions of zero order pressure $p_0$ and energy density $\rho_0$ of holographic fluid, we can easily further obtain $p_0=-\frac{\rho_0(6+\rho_0)}{3(-6+\rho_0)}$. Therefore, the perturbations of energy density and pressure should satisfy the underlying thermodynamic relation through the sonic velocity $c_s$, i.e. $\delta p=c_s^2 \delta\rho$, while the square of sonic velocity can be easily obtain
\begin{eqnarray}
c_s^2=(\frac{\partial p_0}{\partial \rho_0})_{s}=-\frac{\rho_0^2-12\rho_0-36}{3(\rho_0-6)^2}=\frac{(2 M +r_c^4)}{3( -2 M+r_c^4)}, \label{SoundV}
\end{eqnarray}
where the zero order energy density $\rho_0$ and pressure $p_0$ have been used, and the derivative is usually taken for an adiabatic process, i.e. the constant entropy density $s=\frac{r_+^3}{4 G r_c^3}$. In our case, we check that the perturbations of energy density $\delta\rho$ and pressure $\delta p$ should be
\begin{eqnarray}
\delta\rho=-2\theta-\frac{8 \sqrt{f(r_c)}C_{h1}}{r_c^4},~~ \delta p=-\frac{8(2 M+r_c^4)C_{h1}}{3r_c^8\sqrt{f(r_c)}}-\frac{2(2 M +r_c^4)}{3( -2 M+r_c^4)}\theta.
\end{eqnarray}
Therefore, it is obvious and interesting to find that the second choice (\ref{ets1}) will be just the right choice which is consistent to satisfy the underlying thermodynamic relation between the perturbations of energy density and pressure through the sonic velocity, i.e. $\delta p= \frac{(2 M +r_c^4)}{3( -2 M+r_c^4)}\delta\rho=c_s^2 \delta\rho$. In addition, this choice is also consistent with the $T^{(1)}_{v v}=0$ case with zero bulk viscosity in the appendix~\ref{B}. Note that, our proposal of taking the sonic velocity into account also implicates that the true bulk viscosity $\zeta_T$ should be not $\zeta$ but $\zeta_T=\zeta-\zeta^{'}(\frac{\partial p}{\partial \rho})=\zeta-c_s^2 \zeta^{'}$ in (\ref{StressTensor1}), which is underlying consistent with the discussion in \cite{Bhattacharya:2011tra}, where a frame invariant scalar related to the bulk viscosity has been defined in (2.10) and further explicitly obtained in (2.24).

\subsection{Boundary with $h(r_c)\neq0$}
 In the above subsection, we have proposed a method to clarify the superficial similarity between the bulk viscosity and perturbation of the pressure. Note that, during using the Dirichlet boundary condition (\ref{DirichletB}), the main underlying simple reason is to keep a well-defined boosted transformation at the finite cutoff surface $r=r_c$, i.e. $\gamma_{\mu\nu}=\eta_{\mu\nu}$. However, after a careful observation at the corrected metric (\ref{IVcorrection}), we find that the condition $h(r_c)=0$ in (\ref{DirichletB}) in fact can be relaxed as $h(r_c)\neq0$, which also keeps a well-defined boosted transformation at the finite cutoff surface $r=r_c$. The cost is that the traceless condition in (\ref{gauge}) $Tr((G^{(0)})^{-1}G^{(1)})=0$ has been broken as
\begin{equation}
Tr((G^{(0)})^{-1}G^{(1)})=2h(r_c)\ ,\label{newgauge}
\end{equation}
where we have used the deduced condition $\alpha_{xx}(r_c)=\alpha_{yy}(r_c)=\alpha_{zz}(r_c)=\frac{2}{3}h(r_c)$ from the order $\gamma_{\mu\nu}=\eta_{\mu\nu}$. In addition, for the corrected metric in (\ref{IVcorrection}) with a non-traceless $\alpha_{ij}(r)$, i.e. $\sum_i \alpha _{i i}(r)\neq0$, the new components of tensor $W_{AB } = (\text{effect
from correction}) - S_{AB}$ become more complicate, which have also been expressed in Appendix~\ref{C}.

However, from these new components $W_{AB}$, we find that the solutions $h(r)$, $k(r)$ and $j_i(r)$ are same as (\ref{Coefficient}), while $\alpha _{i j}(r)$ can be instead as
\begin{eqnarray}
\alpha_{ij}(r)&=&\alpha(r)\left\{(\partial_i \beta_j + \partial_j
\beta_i )-\frac{2}{3} \delta_{ij}\partial_k \beta^k \right\} { +b\delta_{ij}}, \label{alpha_ijNew}
\end{eqnarray}
where $b$ is a constant. In addition, the first order of stress tensors in (\ref{BoundaryST}) also have been changed and become more complicate
\beqs
    T^{(1)}_{v v} &=&-2\partial _i\beta _i +6 \sqrt{f\left(r_c\right)}h\left(r_c\right)+\frac{\left(-2 C+9 \sqrt{f\left(r_c\right)}\right) k\left(r_c\right)}{ r_c^2}+2\sqrt{f\left(r_c\right)} r_c h'\left(r_c\right){-r_c\sqrt{f\left(r_c\right)}B(r_c)},\nn\\
    T^{(1)}_{v i} &=& \frac{\partial _iM}{f\left(r_c\right) r_c^4}-\partial _v\beta _i +2\frac{\left(2-C \sqrt{f\left(r_c\right)}+3 f\left(r_c\right)\right)  j_i\left(r_c\right)}{\sqrt{f\left(r_c\right)} r_c^2}-\frac{\sqrt{f\left(r_c\right)} j_i'\left(r_c\right)}{ r_c},\nn
\eeqs
\beqs
    T^{(1)}_{i j} &=&2\left( \delta _{i j}\partial _k\beta _k-\partial _{(i}\beta _{j)} \right)+2\delta _{i j}\frac{\partial_v M}{f\left(r_c\right) r_c^4}
 +2\left(-C+\frac{-2 M+3 r_c^4}{\sqrt{f\left(r_c\right)} r_c^4}\right)a_{i j}\left(r_c\right)-\sqrt{f\left(r_c\right)} r_c a_{i j}'\left(r_c\right)\nn\\
    & & { +r_c\sqrt{f\left(r_c\right)}B(r_c)\delta _{i j}}+2\delta _{i j}\left(\left(\frac{2 C }{3}+\frac{5 \left(2 M-3 r_c^4\right)}{3 \sqrt{f\left(r_c\right)} r_c^4}\right) h\left(r_c\right)-\frac{2}{3} \sqrt{f\left(r_c\right)} r_c h'\left(r_c\right)\right)\nn\\
    & & +2\delta _{i j}\left(\frac{\left(-2 M+\left(3-2 f\left(r_c\right)\right) r_c^4\right) k\left(r_c\right)}{2 \sqrt{f\left(r_c\right)} r_c^6}-\frac{\sqrt{f\left(r_c\right)} k'\left(r_c\right)}{2
    r_c}\right).\label{BoundarySTNew}
\eeqs
where $B(r)=\sum_i\alpha _{i i}'(r)$. Therefore, from the Dirichlet boundary condition $k(r_c) = 0,~j_i(r_c) = 0$ and $T^{(1)}_{v x} =0$ in (\ref{DirichletB}) and (\ref{LandauFrame}), we can obtain the parameters $\Cktwo, \Cthree$ and $\Cfour$
\begin{eqnarray}
{\Cktwo}&=&\frac{2 {\Chtwo} r_c^2}{f\left(r_c\right)}-\frac{4 {\Chone} M}{3 r_c^6 f\left(r_c\right)}-\frac{2 r_c^2 \partial _i\beta _i}{3 \sqrt{f\left(r_c\right)}},\nn\\
{\Cthree}&=&-\frac{4 r_c^2 \partial _v\beta _i}{\sqrt{f(r_c)}(2M+r_c^4)},\quad{\Cfour}=\frac{2 M r_c^2 \partial _v\beta _i}{\sqrt{f(r_c)}(2M+r_c^4)} , \label{Param}
\end{eqnarray}
where $\Chone$ and $\Chtwo$  are arbitrary parameters related to the unfixed $h(r_c)$, and $B(r)$ will be found to zero in this case. Substituting (\ref{Param}) into (\ref{BoundarySTNew}), i.e. the non-zero first order stress tensor of holographic fluid at finite cutoff surface, one can obtain
\beqs
    T^{(1)}_{v v} &=&-2\partial _i\beta _i +6 \sqrt{f\left(r_c\right)}{\Chtwo}-\frac{2 \sqrt{f\left(r_c\right)}{\Chone}} {r_c^4} ,\nn\\
    T^{(1)}_{i j} &=& \frac{- 2r_+^3\sigma _{ij}}{r_c^3}+\delta_{ij}\big(\frac{-2(2 M +r_c^4)}{3( -2 M+r_c^4)}\partial _k\beta _k-\frac{2f(r_c) + 12(1-\sqrt{f(r_c)})}{3r_c^4\sqrt{f(r_c)}} C_{h1} \nn\\
   &{}&+\frac{4(1+3\sqrt{f(r_c)})-10f(r_c)}{3\sqrt{f(r_c)}}C_{h2} { -2b(3+\frac{2M-3r_c^4}{r_c^4\sqrt{f(r_c)}})}\big).
\eeqs
Note that, after making some tedious calculations, one can finally obtain a simple result
\beqs
 T^{(1)}_{i j}&=&\frac{- 2r_+^3\sigma _{ij}}{r_c^3}+\delta_{ij}\big(\frac{-2(2 M +r_c^4)}{3( -2 M+r_c^4)}\partial _k\beta _k-\frac{2(2M+r_c^4)}{3 r_c^8 \sqrt{f(r_c)}}C_{h1} +\frac{2(2M+r_c^4)}{r_c^4 \sqrt{f(r_c)}}C_{h2}\big), \nn\\
 &=&\frac{- 2r_+^3\sigma _{ij}}{r_c^3}+\delta_{ij}(c_s^2 T^{(1)}_{v v}),
    \label{NewBoundaryST2}
\eeqs
where the condition $b=\frac{2}{3}h(r_c)$ has been used to keep $\alpha_{xx}(r_c)=\alpha_{yy}(r_c)=\alpha_{zz}(r_c)=\frac{2}{3}h(r_c)$.
From these results and taking the method into account, one will be surprised that transport coefficients can still be precisely extracted although some parameters have not been fixed, i.e. $\Chone$ and $\Chtwo$. We obtain that the bulk viscosity is still zero in this more general boundary condition case with $h(r_c)\neq0$.


\section{Conclusion and discussion}

In this article, after constructing the first order perturbative solution of the Schwarzschild-AdS black brane spacetime, we use the gravity/fluid correspondence to carefully investigate the stress tensor of first order holographic fluid at a finite cutoff surface by considering different boundary conditions. Usually, we select some frame to discuss the fluid, such as Landau frame or Eckart frame in fluid mechanics. However, recent studies show that the physical results may be different in different frames \cite{new}, especially in the studies of stability problem. Therefore, it seems better to relax the constraints of the Landau frame, i.e. admitting the perturbation of energy density in our case. However, an important question is that how we can eliminate the ambiguity freedom in $T^{(1)}_{xx}$ if we relax the constraint. The first key point of our paper is to answer this question, and we obtain that this ambiguity freedom is related to the perturbation of the pressure and bulk viscosity terms in $T^{(1)}_{xx}$, which are very similar. Furthermore, we find a method by taking the sonic velocity in (\ref{SoundV}) into account to clarify this superficial similarity between bulk viscosity and perturbation of the pressure to obtain the physical transport coefficients. The second key point of our paper is that we have explicitly expressed this similarity between bulk viscosity and perturbation of the pressure terms in $T^{(1)}_{xx}$ by investigating another special boundary condition case related to the scalar mode $h(r)$ of metric perturbation, i.e. $h(r_c) =0$ but $h'(r_c)$ is arbitrary, which has not been investigated and seen before, and we find that this condition $h'(r_c)\neq 0$ is in fact crucial to explicitly yield the perturbation of pressure and explicitly see the superficial similarity between pressure perturbation and bulk viscosity. However, by using this method, we can easily obtain the physical transport coefficients in this case. The third key point of our paper is that we further investigate a more general boundary condition case, i.e. $h(r_c)$ is not zero, which has not been considered in previous work yet, too. In this case, one can find that it is more complicate than the cases considered before, since some results have been changed, i.e. the traceless condition $Tr((G^{(0)})^{-1}G^{(1)})$ has been broken and the formula of stress tensor in (\ref{BoundarySTNew}) become more complicate. Moreover, the two parameters $C_{h1}$ and $C_{h2}$ cannot be fixed now due to the nonzero $h(r_c)$. However, it is surprised that one can still extract exact information of transport coefficients from the complicate formula $T^{(1)}_{xx}$ by using the method, and we obtain that bulk viscosity is still zero in this more general boundary condition case.

Note that, our results of sound velocity in holographic fluids via gravity/fluid correspondence are non-trivial. First, our results are the original one among the references because almost of all the previous works via gravity/fluid correspondence are just considered under the Landau frame, i.e. $T^{(1)}_{vv}=0$. It should be pointed out that the corresponding transport coefficients may be also finally obtained under the Landau frame if some boundary condition is lost just like the case with the boundary $h(r_c)\neq 0$ in our paper, but the calculations will be more complicate. Moreover, under the Landau frame, the explicit coefficient $c_s^2$ in front of $T^{(1)}_{v v}$ shall not be obtained in the expression of $T^{(1)}_{xx}$ in (\ref{NewBoundaryST2}). In fact, (\ref{NewBoundaryST2}) is an important equation to show some underlying relationship between $T^{(1)}_{xx}$ and $T^{(1)}_{v v}$. Second, based on the result in the case with the boundary $h(r_c)\neq 0$, it implicates that the sonic velocity can further simplify the complicate expression of $T^{(1)}_{xx}$ in (\ref{BoundaryST}). Indeed, our subsequent work in \cite{Hu:2017vzs} has shown this point. Moreover, after using this simplification, we have further found out an underlying universality in the expression of $T^{(1)}_{xx}$, which has shed some insights on the clue of obtaining the non-zero bulk viscosity for the holographic fluid at the finite cutoff surface. More details and some other discussions related to our work are:

(1) Usually the superficial similarity between bulk viscosity and perturbation of pressure is hardly seen and difficult to distinguish, and we propose an approach to extract the physical transport coefficients of the holographic fluids in our paper. In addition, our proposal also implicates that the true bulk viscosity $\zeta_T$ should be not $\zeta$ but $\zeta_T=\zeta-\zeta^{'}(\frac{\partial p}{\partial \rho})$ in (\ref{StressTensor1}), which is underlying consistent with the discussion in \cite{Bhattacharya:2011tra}, where a frame invariant scalar related to the bulk viscosity has been defined in (2.10) and explicitly obtained in (2.24).

(2) Our approach is useful to find out the true bulk viscosity term in the scenario of gravity/fluid correspondence, and further studies of holographic fluid with other different boundary conditions at the finite cutoff surface are in procedure, where it is indeed very powerful to simplify the $T^{(1)}_{xx}$ by taking the sonic velocity into account in \cite{Hu:2017vzs}. Moreover, we also find some underlying universality in the $T^{(1)}_{xx}$ after taking the sonic velocity into account. In addition, note that here we chose these boundary conditions just simply from the mathematical point, i.e. these boundary conditions are mathematically permitted. However, the underlying physical meaning of these boundary conditions is lost, therefore, it will be interesting and important to find out the underlying physical meaning of these different boundary conditions in the future work. In addition, there have been other methods and works to investigate the bulk viscosity~\cite{Buchel:2007mf,Gubser:2008sz,Yarom:2009mw,Eling:2011ms,Buchel:2011yv,Brattan:2011my,Camps:2010br,Emparan:2012be,Emparan:2013ila}, thus it will be also interesting to make the comparisons between these methods and the method based on gravity/fluid correspondence, which maybe give some insights into the underlying physical meaning of these different boundary conditions, too.

(3) All our discussions are considered in the so-called background gauge in (\ref{IVcorrection}). In fact, as discussed in \cite{Emparan:2013ila}, there is an ambiguity for the extra correction term $g^{(1)}$ in (\ref{IVcorrection}). This ambiguity can affect our choices of the boundary conditions, and hence may affect the stress tensor with transport coefficients. Indeed, there have been several works showing that bulk viscosity can also appear in other gauge~\cite{Emparan:2012be,Emparan:2013ila}. In addition, there are gauge invariant quantities for the metric and energy momentum tensor under perturbation~\cite{Bardeen:1980kt}, and whether bulk viscosity depends on these gauge invariant quantities is still an open issue. Therefore, the underlying relations between gauge, boundary conditions, gauge invariant quantities and stress tensors for holographic fluids with transport coefficients are interesting to be further studied.


\section{Acknowledgements}
Y.P Hu thanks a lot for discussions with Profs. Yan Liu, Ya-Wen Sun, Hai-Qing Zhang, Rong-Gen Cai, Li-Ming Cao and Drs. Song He, Yun-Long Zhang, and also thanks anonymous referees for helpful comments. This work is supported by National Natural Science Foundation of China (NSFC) under grant Nos. 11575083, 11565017, 11105004, the Fundamental Research Funds for the Central Universities under grant No. NS2015073, Shanghai Key Laboratory of Particle Physics and Cosmology under grant No. 11DZ2260700, and the Open Project Program of State Key Laboratory of Theoretical Physics, Institute of Theoretical Physics, Chinese Academy of Sciences, China (No. Y5KF161CJ1). In addition, it is also supported partially by grants from NSFC (No. 10821504, No. 10975168 and No. 11035008), (No. 11205097, No.11175245, No. 11075206 and No.11275128 ), the Ministry of Science
and Technology of China under Grant No. 2010CB833004, the Program for the Innovative Talents of Higher Learning Institutions of Shanxi, the Natural Science Foundation for Young Scientists of Shanxi Province, China (Grant No.2012021003-4), and the Program for Professor of Special Appointment (Eastern Scholar) at Shanghai Institutions of Higher Learning. In addition, Y.P Hu thanks a lot for the support from the Sino-Dutch scholarship programme under the CSC scholarship.

\appendix
\section{The tensor components of $W_{AB }$ and $S_{AB }$}
\label{A} The tensor components of $W_{AB } = (\text{effect
from correction}) - S_{AB}$ are
\begin{eqnarray}
& &W_{vv} =-\frac{8 r^2 f(r) h(r)}{r_c^2 f\left(r_c\right)}-\frac{2 \left(2 M+r^4\right) f(r) h'(r)}{r r_c^2 f\left(r_c\right)}+\frac{f(r) k'(r)}{2 r}-\frac{1}{2} f(r) k''(r)-S_{vv}^{(1)}~,\nn\\
& &W_{vi}=\frac{3 f(r) j_i'(r)}{2 r}-\frac{1}{2} f(r) j_i''(r)-S_{vi}^{(1)}(r)~,\\
& &W_{vr}=\frac{8 h(r)}{r_c \sqrt{f\left(r_c\right)}}+\frac{2 \left(2 M +r^4\right) h'(r)}{r^3 r_c \sqrt{f\left(r_c\right)}}-\frac{r_c \sqrt{f\left(r_c\right)} k'(r)}{2 r^3}+\frac{r_c \sqrt{f\left(r_c\right)} k''(r)}{2 r^2}-S_{vr}^{(1)},
\end{eqnarray}
\begin{eqnarray}
& &W_{ri} =-\frac{3 r_c \sqrt{f\left(r_c\right)} j_i'(r)}{2 r^3}+\frac{r_c \sqrt{f\left(r_c\right)} j_i''(r)}{2 r^2}-S_{ri}^{(1)},\\
& &W_{rr} = \frac{5 h'(r)}{r}+h''(r)-S_{rr}^{(1)},\\
& &W_{ii}=\frac{8 r^2}{r_c^2} h(r)+\frac{\left(-14 M +11 r^4\right) h'(r)}{3 r r_c^2}+\frac{1}{3r_c^2} r^4 f(r) h''(r)+\frac{f\left(r_c\right) k'(r)}{r}\nn\\
& &\qquad+\frac{\left(2 M -5 r^4\right) \alpha _{i i}'(r)}{2 r r_c^2}-\frac{1}{2 r_c^2} r^4 f(r) \alpha _{i i}''(r)-S_{ii}^{(1)}, (\text{here $ii=xx, yy, zz$ with no summation}) \label{A6}\\
& &W_{ij}=\frac{\left(2 M -5 r^4\right) \alpha _{i j}'(r)}{2 r r_c^2}-\frac{1}{2r_c^2} r^4 f(r) \alpha _{i j}''(r)-S_{ij}^{(1)},~(i\neq j),\\
& &W_{ij}-\dfrac{1}{3}\delta_{ij}\left(\sum_k W_{kk}\right)=\frac{\left(2 M -5 r^4\right) \alpha _{i j}'(r)}{2 r r_c^2}-\frac{1}{2r_c^2} r^4 f(r) \alpha _{i j}''(r)-S_{ij}^{(1)}+\dfrac{1}{3}\delta_{ij}(\delta^{kl}S_{kl}^{(1)}), \label{A8}
\end{eqnarray}
where the first order source terms are
\begin{eqnarray}
S_{vv}^{(1)}(r)&=&-\frac{3\partial _vM}{r^3 r_c \sqrt{f\left(r_c\right)}}-\frac{\left(2 M +r^4\right) \partial _i\beta _i}{r^3 r_c \sqrt{f\left(r_c\right)}},\\
S_{vi}^{(1)}(r)&=&\frac{\left(-2 M +3 r^4+2 r_c^4\right) \partial _iM}{2 r^3 r_c^5 f\left(r_c\right){}^{3/2}}+\frac{\left(2 M +3 r^4\right) \partial _v\beta _i}{2 r^3 r_c \sqrt{f\left(r_c\right)}},\\
S_{vr}^{(1)}(r)&=&\frac{\partial _i\beta _i}{r},\\
S_{ri}^{(1)}(r)&=&-\frac{3 \partial _v\beta _i}{2 r}-\frac{3 \partial _iM}{2 r r_c^4 f\left(r_c\right)},\\
S_{rr}^{(1)}(r) &=&0, \\
S_{ij}^{(1)}(r) &=&\left(\delta _{i j}\partial _k\beta _k+3\partial _{(i}\beta _{j)}\right)\frac{r\sqrt{f(r_c)}}{r_c}.
\end{eqnarray}
\section{The case $T^{(1)}_{v v} =0$} \label{B}
In this case, the nine parameters can be fixed from $T^{(1)}_{v v} =0$, (\ref{DirichletB}) and (\ref{LandauFrame}) \beqs
    & &{\Chone}=-\frac{\partial _i\beta _i r_c^4}{4 \sqrt{f\left(r_c\right)}},\quad{\Chtwo}=\frac{\partial _i\beta _i}{4 \sqrt{f\left(r_c\right)} },~~\quad{\Cktwo}=-\frac{\partial _i\beta _i \left(-10 M+r_c^4\right)}{6 f\left(r_c\right){}^{3/2} r_c^2},\nn\\
    & &{\Cthree}=-\frac{4 r_c^2 \partial _v\beta _i}{\sqrt{f(r_c)}(2M+r_c^4)},\quad{\Cfour}=\frac{2 M r_c^2 \partial _v\beta _i}{\sqrt{f(r_c)}(2M+r_c^4)}\ . \label{parameters}
\eeqs

Consequently, the non-zero components of $T^{(1)}_{\mu\nu}$ are
\beqs
   & &T^{(1)}_{i j} = - 2r_+^3\sigma _{ij}/r_c^3\ ,~~\sigma_{ij}=\partial_{(i} \beta_{j)}-\frac{1}{3} \delta_{ij}\partial_k \beta^k. \label{APixBoundaryST}
\eeqs

From (\ref{StressTensor1}), one can simply read out
\begin{eqnarray}
\rho=6\left(1- \sqrt{f\left(r_c\right)} \right),~~~p=\frac{-4
M+6\left(1- \sqrt{f\left(r_c\right)}\right)
r_c^4}{r_c^4\sqrt{f\left(r_c\right)}}, ~~~\eta=r_+^3/r_c^3,~~\zeta=0. \label{IVets}
\end{eqnarray}
Thus, the dual fluid obtained at
the finite cutoff surface is indeed not conformal because the trace of $T_{\mu\nu}$ is nonzero, i.e. $\rho=3p$ has been broken. This result is consistent with that in Ref.~\cite{Kuperstein:2011fn}, and expected from the fact that the conformal symmetry has been broken with a finite
radial coordinate in the bulk. In addition, as $r_c \rightarrow \infty$, the results in (\ref{IVets}) can relate to those in the infinite boundary by just a conformal factor. Since the conformal symmetry is recovered at this case, these results can be related to each other by conformal transformation. Moreover, since the entropy density from (\ref{IVSolution2}) is $s=\frac{r_+^3}{4 G r_c^3}$, and after recovering the coefficient $16\pi G$ in $\eta$, we can easily find that $\eta/s=1/(4 \pi)$, which is consistent with the well-known $\eta/s$ result for the dual fluid at the infinite boundary in the Einstein gravity~\cite{Bhattacharyya:2008jc,Hur:2008tq,Hu:2010sn, Hu:2011ze,Bai:2012ci}.

\section{New tensor components of $W_{AB }$ and $S_{AB }$}
\label{C}
For the corrected metric in (\ref{IVcorrection}) with a non-traceless $\alpha_{ij}(r)$ i.e., $\sum_i \alpha _{i i}(r)\neq0$, we can obtain the new tensor components of $W_{AB } = (\text{effect
from correction}) - S_{AB}$ are
\begin{eqnarray}
& &W_{vv} =-\frac{8 r^2 f(r) h(r)}{r_c^2 f\left(r_c\right)}-\frac{2 \left(2 M+r^4\right) f(r) h'(r)}{r r_c^2 f\left(r_c\right)}+\frac{f(r) k'(r)}{2 r}-\frac{1}{2} f(r) k''(r)+\frac{(2M+r^4)r_c^2(2M-r^4)}{2r^5(2M-r_c^4)}B(r)-S_{vv}^{(1)}~,\nn\\
& &W_{vi}=\frac{3 f(r) j_i'(r)}{2 r}-\frac{1}{2} f(r) j_i''(r)-S_{vi}^{(1)}(r)~,\nn\\
& &W_{vr}=\frac{8 h(r)}{r_c \sqrt{f\left(r_c\right)}}+\frac{2 \left(2 M +r^4\right) h'(r)}{r^3 r_c \sqrt{f\left(r_c\right)}}-\frac{r_c \sqrt{f\left(r_c\right)} k'(r)}{2 r^3}+\frac{r_c \sqrt{f\left(r_c\right)} k''(r)}{2 r^2}-\frac{2M+r^4}{2r^3r_c\sqrt{f(r_c)}}B(r)-S_{vr}^{(1)},\nn\\
& &W_{ri} =-\frac{3 r_c \sqrt{f\left(r_c\right)} j_i'(r)}{2 r^3}+\frac{r_c \sqrt{f\left(r_c\right)} j_i''(r)}{2 r^2}-S_{ri}^{(1)},\nn\\
& &W_{rr} = \frac{5 h'(r)}{r}+h''(r)-\frac{B(r)}{r}-\frac{B'(r)}{2}-S_{rr}^{(1)},\nn\\
& &W_{ii}=\frac{8 r^2}{r_c^2} h(r)+\frac{\left(-14 M +11 r^4\right) h'(r)}{3 r r_c^2}+\frac{1}{3r_c^2} r^4 f(r) h''(r)+\frac{f\left(r_c\right) k'(r)}{r}\nn\\
& &\qquad+\frac{\left(2 M -5 r^4\right) \alpha _{i i}'(r)}{2 r r_c^2}-\frac{1}{2 r_c^2} r^4 f(r) \alpha _{i i}''(r)-\frac{r^3f(r)}{2r_c^2}B(r)-S_{ii}^{(1)}, (\text{here $ii=xx, yy, zz$ with no summation}) \nn\\
& &W_{ij}=\frac{\left(2 M -5 r^4\right) \alpha _{i j}'(r)}{2 r r_c^2}-\frac{1}{2r_c^2} r^4 f(r) \alpha _{i j}''(r)-S_{ij}^{(1)},~(i\neq j),\nn\\
& &W_{ij}-\dfrac{1}{3}\delta_{ij}\left(\sum_k W_{kk}\right)=\frac{\left(2 M -5 r^4\right) \left(\alpha _{i j}'(r)- \delta_{ij} \frac{1}{3} B(r)\right)}{2 r r_c^2} -\frac{1}{2r_c^2} r^4 f(r) \left(\alpha _{i j}'(r) +\delta_{ij} \frac{1}{3} B(r)\right)'-S_{ij}^{(1)}+\dfrac{1}{3}\delta_{ij}(\delta^{kl}S_{kl}^{(1)}),\nn
\end{eqnarray}
where $B(r)=\sum_i\alpha _{i i}'(r)$, and the first order source terms are same as those in Appendix~\ref{A}.

\end{document}